\documentclass[aps,prb,superscriptaddress,amssym,floatfix,notitlepage]{revtex4-1}
\usepackage{amsfonts}
\usepackage{amsmath}
\usepackage{amssymb}
\usepackage{graphicx}

\begin{document}

\title{Effect of long-range interaction on graphene edge magnetism}
\author{Zheng Shi}
\author{Ian Affleck}
\affiliation{Department of Physics and Astronomy, University of British Columbia,
Vancouver, BC V6T 1Z1, Canada}
\date{\today }

\begin{abstract}
It has been proposed that interactions lead to ferromagnetism on a zigzag edge of a graphene sheet. While not yet directly studied experimentally, dramatically improving techniques for making and studying clean zigzag edges may soon make this possible. So far, most theoretical investigations of this claim have been based on mean field theories or more exact calculations using the Hubbard model. But long-range Coulomb interactions are unscreened in graphene so it is important to consider their effects. We study rather general non-local interactions, including of Coulomb $1/r$ form, using the technique of projection to a strongly interacting edge Hamiltonian, valid at first order in the interactions. The ground states as well as electron/hole and exciton excitations are studied in this model. Our results indicate that ferromagnetism survives with unscreened Coulomb interactions.
\end{abstract}

\maketitle

\section{Introduction\label{sec:intro}}

Non-interacting graphene nanoribbons with zigzag edges are famous for
hosting a nearly flat band of edge states.\cite%
{JPSJ.65.1920,1468-6996-11-5-054504} In the presence of electron-electron
interaction, the existence of edge magnetic order\cite{0034-4885-73-5-056501}
has been predicted by a multitude of theoretical work using both analytical%
\cite%
{JPSJ.65.1920,PhysRevB.68.193410,PhysRevLett.102.227205,*PhysRevB.79.235433,PhysRevB.80.155441,PhysRevB.82.085422,PhysRevB.83.165415,1367-2630-13-3-033028,PhysRevB.86.075458,PhysRevB.86.115446}
and numerical\cite%
{PhysRevB.83.195432,PhysRevB.87.245431,PhysRevLett.101.036803,PhysRevB.87.155441,PhysRevLett.111.085504,PhysRevLett.112.046601,*PhysRevB.87.245431,PhysRevB.94.165147,PhysRevLett.97.216803,PhysRevB.75.064418,0953-8984-25-5-055304}
techniques. The consensus emerging from these work is that edge states
localized at the same edge are coupled ferromagnetically to form superspins,
which then couple antiferromagnetically between edges. In addition to ground
state properties, low-energy magnetic excitations in graphene nanoribbons
have also attracted much theoretical attention.\cite%
{JPSJ.67.2089,PhysRevLett.100.047209,PhysRevB.80.155441,1367-2630-13-3-033028}
A relatively large spin correlation length up to the order of micrometers
has been found for a single zigzag edge; this is attributed to the large
spin stiffness in this system, and boosts confidence in potential
spintronics applications of graphene edge magnetism.\cite{PhysRevB.81.165404}
Although conclusive experimental evidence for edge magnetism is still
lacking due to limited control over edge orientation, there has been
significant progress in recent years towards the synthesis and
characterization of zigzag edges.\cite%
{Nature.514.608,Nature.531.489,SciRep.5.13382}

A large number of theoretical studies on graphene edge magnetism represent
the interaction by an on-site Hubbard term for simplicity. For the Hubbard
model on a honeycomb lattice, arguments in support of edge magnetism\cite%
{PhysRevB.86.115446} can be constructed based on Lieb's theorem.\cite%
{PhysRevLett.62.1201} The Coulomb interaction in pristine graphene on a
non-metallic substrate is, nevertheless, poorly screened due to a vanishing
density of states at the Dirac points.\cite%
{PhysRevB.29.1685,RevModPhys.84.1067} The influence of non-local components
of the interaction has been investigated both in bulk graphene\cite%
{PhysRevB.89.205128,PhysRevLett.97.146401,PhysRevLett.100.146404,PhysRevLett.111.036601}
and in restricted geometries.\cite%
{PhysRevB.68.193410,PhysRevB.83.165415,PhysRevB.90.155433,0953-8984-18-27-010,PhysRevLett.101.036803} (By ``non-local'' we mean having a longer range than on-site.)
However, many studies on graphene nanoribbons with non-local interactions
have adopted a mean-field treatment, neglecting fluctuations whose role is
especially important in low dimensions.\cite{PhysRevLett.17.1133} Exact
diagonalization has been employed in other studies; despite the light it
sheds on the nature of the ground states, correlations in manageably small
systems are usually enhanced compared to the thermodynamic limit.

In the present work, we study the effect of long-range interactions on graphene
edge ferromagnetism, in the limit of weak interactions but beyond the
mean-field level. Focusing on a semi-infinite graphene sheet with a single
zigzag edge, we find the effective Hamiltonian by projecting the interaction
into the Hilbert space of edge states; we then propose a sufficient
condition for the maximum spin ferromagnetic multiplet to be the
half-filling ground states. Using exact diagonalization, we discuss the
possible ground states for interactions in violation of this condition. The
long-range Coulomb interaction is shown to satisfy the sufficient condition
upon extrapolation to the limit of infinite long distance cutoff. We also
examine the simplest low-energy excitations of the ferromagnetic ground
states on a single edge. For short-range interactions, single-particle
excitations and single-hole excitations have linear spectra $\propto v\delta
k$ where $\left\vert \delta k\right\vert \ll 1$ is the distance from either
Dirac point, with a slope $v$ controlled by the interaction strength. Spin-1
excitons have a small-momentum dispersion that is proportional to $vQ^{2}\ln
Q$. For the long-range Coulomb interaction, $v\rightarrow \infty $, and the
dispersion of single-particle or single-hole excitations near the Dirac
points scales as $\delta k\ln \delta k$. Finally, for both short-range and
Coulomb interactions, a sufficiently large particle-hole symmetry breaking
term in the Hamiltonian can destabilize the ferromagnetic ground state.

\section{Model\label{sec:model}}

\begin{figure}[ptb]
\includegraphics[width=0.6\textwidth]{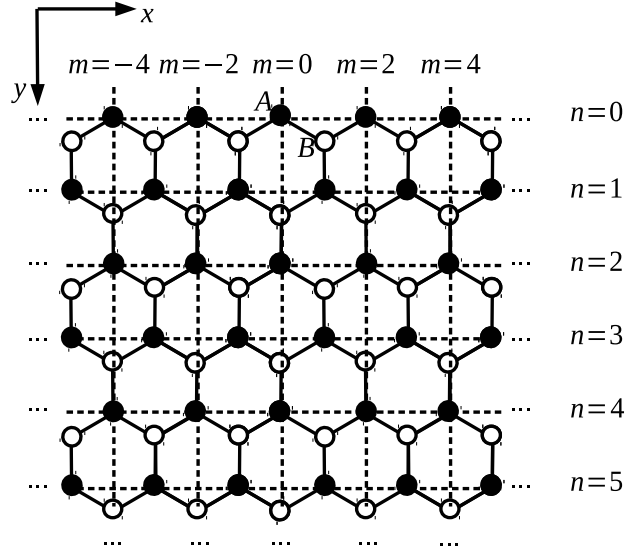}%
\caption{Sketch of a
semi-infinite graphene sheet with a zigzag edge.\label{fig:sketch}}
\end{figure}

We study a semi-infinite graphene sheet on the $xy$ plane, modeled by a
honeycomb lattice which is terminated by an infinite zigzag edge (see Fig.~%
\ref{fig:sketch}). All carbon atoms reside in the half plane $y\geq 0$, and
the outermost atoms on the zigzag edge (which belong to the $A$ hexagonal
sublattice) lie on the $x$ axis. In units of the Bravais lattice constant $%
a=2.46$\AA , it is convenient to represent the position of carbon atoms by\ $%
\vec{r}_{\left( m,n\right) }=\left( m/2\right) \hat{x}+\left( \sqrt{3}%
n/2\right) \hat{y}$ where $n\geq 0$. While $m$ is always an integer, note
that $n$ is an integer only on the $A$ sublattice:\ for $A$ atoms $n$ and $m$
are both even or both odd, while for $B$ atoms $n+2/3$ and $m$ are both even
or both odd.

The zero modes associated with the zigzag edge are given by\cite%
{JPSJ.65.1920,1468-6996-11-5-054504}

\begin{equation}
e_{k}^{\dag }=\frac{1}{\sqrt{2\pi }}\sum_{n\geq 0,m}e^{ik\frac{m}{2}%
}g_{n}\left( k\right) c_{m,n,A}^{\dag }\text{,}  \label{edag}
\end{equation}%
where $k$ is the crystal momentum along the edge direction,

\begin{equation}
g_{n}\left( k\right) \equiv \theta \left( k-\frac{2\pi }{3}\right) \theta
\left( \frac{4\pi }{3}-k\right) \sqrt{1-4\cos ^{2}\frac{k}{2}}\left( -2\cos 
\frac{k}{2}\right) ^{n}
\end{equation}%
describes the decay of the wave function into the bulk, and the $c$
operators obey the usual anticommutation relations $\left\{
c_{m,n,A},c_{m^{\prime },n^{\prime },A}^{\dag }\right\} =\left\{
c_{m,n,B},c_{m^{\prime },n^{\prime },B}^{\dag }\right\} =\delta _{mm^{\prime
}}\delta _{nn^{\prime }}$, $\left\{ c_{m,n,A},c_{m^{\prime },n^{\prime
},B}^{\dag }\right\} =0$. (We have temporarily suppressed the spin index.)
These edge states exist only for $2\pi /3<k<4\pi /3$, i.e. in $1/3$ of the
1D Brillouin zone $0\leq k<2\pi $. The wave function is non-zero only on the 
$A$ sublattice, and is localized near the zigzag edge. The localization
length $\xi _{k}=-\left[ \ln \left\vert 2\cos \left( k/2\right) \right\vert %
\right] ^{-1}$ vanishes at $k=\pi $, and diverges near the Dirac points $%
k=2\pi /3$ and $k=4\pi /3$.

In addition to the edge states, we also have bulk states which are labeled
by $k$, $k_{y}$ and $s$:

\begin{eqnarray}
b_{k,k_{y},s}^{\dag } &=&\frac{1}{2\pi }\frac{1}{\sqrt{2}}\left\{
\sum_{n\geq 0,m}e^{ik\frac{m}{2}}\left[ 2i\sin nk_{y}+\left( 2\cos \frac{k}{2%
}\right) 2i\sin \left( n+1\right) k_{y}\right] \right.   \notag \\
&&\left. \times \frac{t}{E_{s}\left( k,k_{y}\right) }c_{m,n,A}^{\dag
}-\sum_{n\geq \frac{1}{3},m}e^{ik\frac{m}{2}}\left[ 2i\sin \left( n+\frac{2}{%
3}\right) k_{y}\right] c_{m,n,B}^{\dag }\right\} \text{.}  \label{bdag}
\end{eqnarray}%
Here the bulk dispersion relation is

\begin{equation}
E_{s}\left( k,k_{y}\right) =st\sqrt{\left( 2\cos \frac{k}{2}\right)
^{2}+1+2\left( 2\cos \frac{k}{2}\right) \cos k_{y}}
\end{equation}%
with nearest neighbor hopping strength $t$; $k_{y}$ is the crystal
momentum perpendicular to the edge, $0\leq k_{y}\leq \pi $, and $s=\pm $ is
a subband index. Near the Dirac points, where $\left( k,k_{y}\right) =\left(
2\pi /3,\pi \right) +\left( \delta k,\delta k_{y}\right) $ or $\left(
k,k_{y}\right) =\left( 4\pi /3,0\right) +\left( \delta k,\delta k_{y}\right) 
$, $E_{s}\left( k,k_{y}\right) $ takes a Lorentz invariant form $E_{s}\left(
k,k_{y}\right) =st\sqrt{\left( \delta k_{y}\right) ^{2}+\left( 3/4\right)
\left( \delta k\right) ^{2}}$. In this non-interacting model, at zero
temperature and half-filling, the $s=-$ subband is completely filled and the 
$s=+$ subband is completely empty. While the edge states are half-filled,
for the semi-infinite sheet we cannot ascertain which half is filled at this
point, unless other ingredients--such as next-nearest-neighbor hopping, edge
potential and electron-electron interaction--are present.

We now introduce a weak repulsive electron-electron interaction. The
following extended Hubbard model manifestly respects $SU\left( 2\right) $
spin symmetry, and also particle-hole symmetry at half-filling:

\begin{equation}
H_{int}=\frac{1}{2}\sum_{n,m}\sum_{\delta _{m},\delta _{n}}U_{\left( \delta
_{m},\delta _{n}\right) }\left( \sum_{\sigma =\pm }c_{m,n,\sigma }^{\dag
}c_{m,n,\sigma }-1\right) \left( \sum_{\sigma ^{\prime }=\pm }c_{m+\delta
_{m},n+\delta _{n},\sigma ^{\prime }}c_{m+\delta _{m},n+\delta _{n},\sigma
^{\prime }}-1\right) \text{.}  \label{Hint}
\end{equation}%
$\left( \delta _{m},\delta _{n}\right) $ runs over all vectors $\vec{\delta}%
=\left( \delta _{m}/2\right) \hat{x}+\left( \sqrt{3}\delta _{n}/2\right) 
\hat{y}$ pointing from one lattice site to another; for instance, $U_{\left(
0,0\right) }$ stands for the strength of the on-site Hubbard\ interaction, $%
U_{\left( 0,2/3\right) }$ is the interaction between nearest neighbor sites
(belonging to different sublattices) in the $y$ direction, $U_{\left(
1,1/3\right) }$ is the interaction between nearest neighbor sites at $\pi /6$
angle with the $x$ direction, and $U_{\left( 2,0\right) }$ is the
interaction between next nearest neighbors (belonging to the same
sublattice) in the $x$ direction. The sum over $n$ and $\delta _{n}$ is such
that both $n\geq 0$ and $n+\delta _{n}\geq 0$. To lighten notations, we
have suppressed the sublattice indices $A$ and $B$ in this expression,
because they are uniquely determined by the position indices $\left(
m,n\right) $ and $\left( m+\delta _{m},n+\delta _{n}\right) $.

In general $U_{\left( \delta _{m},\delta _{n}\right) }=U_{\left( -\delta
_{m},-\delta _{n}\right) }$, but apart from this constraint $U$ can be an
arbitrary function of $\delta _{m}$ and $\delta _{n}$. Nevertheless we
further assume that $U$ obeys\ parity symmetry, $U_{\left( \delta
_{m},\delta _{n}\right) }=U_{\left( -\delta _{m},\delta _{n}\right) }$. For
the Hubbard model, $U_{\left( \delta _{m},\delta _{n}\right) }$ vanishes
unless $\delta _{m}=\delta _{n}=0$. On the other hand, for the unscreened
Coulomb interaction, $U_{\left( \delta _{m},\delta _{n}\right) }$ is
inversely proportional to distance at large distances,\cite%
{PhysRevLett.79.5082,PhysRevLett.99.256804,*PhysRevLett.101.196804,PhysRevB.83.165415}

\begin{equation}
U_{\left( \delta _{m},\delta _{n}\right) }=U_{0}\frac{d}{\sqrt{%
d^{2}+\left\vert \vec{\delta}\right\vert ^{2}}}\text{,}  \label{CoulombU}
\end{equation}%
where $U_{0}$ is the on-site interaction, and the half-nearest-neighbor
distance $d=1/\left( 2\sqrt{3}\right) $ accounts for the finite spread of
the carbon $\pi $ orbitals.

Assuming $U_{\left( \delta _{m},\delta _{n}\right) }\ll t$, we expect that
the low-energy degrees of freedom are composed of the edge states $e_{k}$
with $2\pi /3<k<4\pi /3$,\ and the bulk states in the vicinity of the two
Dirac points.\cite{PhysRevLett.101.036803,PhysRevB.87.245431} As a first
approximation at $O\left( U\right) $, we neglect the dynamics of the bulk
states completely; they are assumed to be half-filled and not spin-polarized
as in the non-interacting case.\cite{PhysRevB.82.085422,PhysRevB.86.115446}
This approximation allows the projection of the interaction onto the Hilbert
space of the edge states. More concretely, we invert Eqs.~(\ref{edag}) and (%
\ref{bdag}) to express the $c$ operators in terms of $e$ and $b$, then take
the expectation values for pairs of $b$ operators using

\begin{equation}
\left\langle b_{k,k_{y},-,\sigma }^{\dag }b_{k^{\prime },k_{y}^{\prime
},-,\sigma ^{\prime }}\right\rangle =\left\langle b_{k,k_{y},+,\sigma
}b_{k^{\prime },k_{y}^{\prime },+,\sigma ^{\prime }}^{\dag }\right\rangle
=\delta _{\sigma \sigma ^{\prime }}\delta \left( k-k^{\prime }\right) \delta
\left( k_{y}-k_{y}^{\prime }\right) \text{.}
\end{equation}%
After some algebra, we find

\begin{equation}
H_{\text{int}}=\frac{1}{2}\sum_{n}\sum_{\delta _{m},\delta _{n}}U_{\left(
\delta _{m},\delta _{n}\right) }\int_{-\frac{2\pi }{3}}^{\frac{2\pi }{3}}%
\frac{dq}{2\pi }e^{iq\frac{\delta _{m}}{2}}O_{n+\delta _{n}}^{\dag }\left(
q\right) O_{n}\left( q\right) \text{,}  \label{Hintproj}
\end{equation}%
where the sum over $\left( \delta _{m},\delta _{n}\right) $ is now limited
to vectors on one of the sublattices; recalling that edge states only exist
on the $A$ sublattice, $\delta _{m}$ and $\delta _{n}$ are now both even or
both odd. Again $n\geq 0$ and $n+\delta _{n}\geq 0$. $O_{n}\left( q\right) $ is bilinear in $e$,

\begin{equation}
O_{n}\left( q\right) \equiv \int dkg_{n}\left( k+q\right) g_{n}\left(
k\right) \left[ \sum_{\sigma =\pm }e_{k+q,\sigma }^{\dag }e_{k,\sigma
}-\delta \left( q\right) \right] \text{.}
\end{equation}%
$q$ measures the momentum difference between two edge states, so the
operator $O_{n}\left( q\right) $ is nontrivial only when $\left\vert
q\right\vert <2\pi /3$. Note that $O_{n}\left( q\right) $ annihilates all
members of the fully polarized ferromagnetic multiplet at half-filling for
any $n$ and $q$, which means the ferromagnetic multiplet states are always
eigenstates of $H_{\text{int}}$ with zero energy.

Due to the constraint on the $\left( \delta _{m},\delta _{n}\right) $
summation, many terms in the interaction (most notably the nearest-neighbor
interaction) do not enter the projected effective Hamiltonian in the edge
state subspace, Eq.~(\ref{Hintproj}). Although the authors of Ref.~%
\onlinecite{PhysRevB.68.193410} predict a charge-polarized ground state when
the nearest-neighbor interaction prevails over the on-site interaction, our
picture is consistent with their weak interaction limit, where the
charge-polarized state always has a higher energy and the nearest-neighbor
interaction is unimportant.

Just as Eq.~(\ref{Hint}), Eq.~(\ref{Hintproj}) manifestly respects $SU\left(
2\right) $ symmetry and particle-hole symmetry at half-filling. In
particular, the particle-hole transformation $c_{m,n,\sigma }\rightarrow
c_{m,n,\sigma }^{\dag }$ corresponds to $e_{k,\sigma }\rightarrow e_{2\pi
-k,\sigma }^{\dag }$ and $O_{n}\left( q\right) \rightarrow -O_{n}\left(
q\right) $ in the edge state subspace. (The form $e_{k,\sigma }\rightarrow
e_{k,\sigma }^{\dag }$ previously suggested in the Hubbard model\cite%
{PhysRevB.86.115446} is the combination of a particle-hole transformation
and a parity transformation.) The particle-hole symmetry is broken by either
a weak next-nearest neighbor hopping $\left\vert t_{2}\right\vert \ll t$\ in
the bulk, or a weak potential localized at the edge $\left\vert
V_{e}\right\vert \ll t$; the latter can arise, for example, at a
graphene-graphane interface.\cite{PhysRevB.82.085422} When $\Delta
=t_{2}-V_{e}\neq 0$, a dispersion develops for the edge states:

\begin{equation}
H=H_{\text{int}}+H_{\Delta }\text{, }H_{\Delta }=\Delta \sum_{\sigma =\pm
}\int_{\frac{2\pi }{3}}^{\frac{4\pi }{3}}dk\left( 2\cos k+1\right)
e_{k,\sigma }^{\dag }e_{k,\sigma }\text{,}  \label{Hproj}
\end{equation}%
assuming the Fermi energy is fixed at the new Dirac point $\epsilon
_{F}=3t_{2}$.\cite{PhysRevB.82.085422,PhysRevB.86.115446}

In the remainer of this paper we analyze the edge state Hamiltonian given by
Eq.~(\ref{Hproj}) at half-filling.

\section{Ground state\label{sec:gs}}

We first study the ground state of the particle-hole symmetric Hamiltonian
Eq.~(\ref{Hintproj}), keeping $\Delta =0$.

For the projected Hubbard model, it has been proven in Ref.~%
\onlinecite{PhysRevB.86.115446} that the fully polarized ferromagnetic
multiplet states with maximum total spin are the unique ground states. In
the Hubbard case, Eq.~(\ref{Hintproj}) becomes

\begin{equation}
H_{\text{int, Hubbard}}=\frac{1}{2}U\sum_{n=0}^{\infty}\int_{-\frac{2\pi }{3}}^{\frac{%
2\pi }{3}}\frac{dq}{2\pi }O_{n}^{\dag }\left( q\right) O_{n}\left( q\right) 
\text{;}
\end{equation}%
It is obvious that $H_{\text{int, Hubbard}}$ is positive semi-definite.
Since the ferromagnetic multiplet states are always zero energy eigenstates,
they must belong to the ground state manifold of $H_{\text{int, Hubbard}}$.
Furthermore, it is also possible to show that they are the only states
annihilated by $O_{n}\left( q\right) $ for any $n$ and $q$, and therefore
the unique ground states of $H_{\text{int, Hubbard}}$.\cite%
{PhysRevB.86.115446} We emphasize again that the proof rests on the positive
semi-definiteness of the Hamiltonian.

Let us explore the extent to which the proof outlined above can be
generalized in our extended Hubbard model. In analogy to a semi-infinite tight-binding chain, through the following
transformation

\begin{equation}
O_{n}\left( q\right) =\int_{0}^{\pi }\frac{dK}{\pi }O_{K}\left( q\right)
\sin K\left( n+1\right) \text{,}
\end{equation}%
the generic interaction Hamiltonian Eq.~(\ref{Hintproj}) can be formally
diagonalized:

\begin{equation}
H_{\text{int}}=\frac{1}{2}\int_{-\frac{2\pi }{3}}^{\frac{2\pi }{3}}\frac{dq}{%
2\pi }\int_{0}^{\pi }\frac{dK}{2\pi }\tilde{U}\left( k,q\right) O_{K}^{\dag
}\left( q\right) O_{K}\left( q\right) \text{,}
\end{equation}%
where

\begin{equation}
\tilde{U}\left( K,q\right) \equiv \sum_{\delta _{m},\delta _{n}}U_{\left(
\delta _{m},\delta _{n}\right) }\cos \left( K\delta _{n}\right) \cos \frac{%
q\delta _{m}}{2}\text{.}
\end{equation}%
The spectrum of $\tilde{U}\left( K,q\right) $ does not give the spectrum of
the interacting problem because $O_{K}\left( q\right) $ does not obey simple
commutation relations. Nevertheless, if $\tilde{U}\left( K,q\right) $ is
positive semi-definite for $0\leq K\leq \pi $ and $-2\pi /3\leq q\leq 2\pi /3
$, we can borrow the arguments from the case of the Hubbard model, and show
that the ferromagnetic multiplet states are the unique ground states of Eq.~(%
\ref{Hintproj}) at half filling. [That a state is annihilated by all $%
O_{n}\left( q\right) $ is equivalent to it being annihilated by all $%
O_{K}\left( q\right) $.] The positive semi-definiteness of $\tilde{U}\left(
K,q\right) $ is thus a sufficient condition for ferromagnetic ground states.

As a simple example, we consider the model with only on-site and
next-nearest-neighbor interactions:

\begin{equation}
U_{\left( 0,0\right) }\equiv U\text{, }U_{\left( \pm 2,0\right) }\equiv
U_{2\parallel }\text{, }U_{\left( \pm 1,1\right) }=U_{\left( \pm 1,-1\right)
}\equiv U_{2\angle }\text{,}  \label{HintNNN}
\end{equation}%
and $U_{\left( \delta _{m},\delta _{n}\right) }=0$ for other $\left( \delta
_{m},\delta _{n}\right) $. (The nearest-neighbor interactions drop out, as
remarked in Section~\ref{sec:model}.) In the next-nearest-neighbor
interaction we have introduced an anisotropy between the direction parallel
to the edge ($U_{2\parallel }$) and the directions at an angle of $\pi /3$
with the edge ($U_{2\angle }$). While such anisotropy is not necessarily
realistic, we shall see that $U_{2\parallel }$ and $U_{2\angle }$ have very
different effects on edge magnetism.

For this model,

\begin{equation}
\tilde{U}\left( K,q\right) =U+2U_{2\parallel }\cos q+4U_{2\angle }\cos K\cos 
\frac{q}{2}\text{;}
\end{equation}%
as $\cos q/2>0$, the minimum of $\tilde{U}$ with respect to $K$ is obtained
at $K=\pi $. The positive semi-definiteness condition of $\tilde{U}\left(
K,q\right) $ is therefore equivalent to

\begin{equation}
\forall q\in \left[ -\frac{2\pi }{3},\frac{2\pi }{3}\right] \text{, }%
U+2U_{2\parallel }\cos q\geq 4U_{2\angle }\cos \frac{q}{2}\text{.}
\label{NNNFM}
\end{equation}%
This is a sufficient condition for the ground states to be ferromagnetic in
the model specified by Eq.~(\ref{HintNNN}). It requires that neither $%
U_{2\parallel }$ nor $U_{2\angle }$ should be greater than $U$. In
particular, Eq.~(\ref{NNNFM}) becomes $U_{2\angle }\leq U/4$ when $%
U_{2\parallel }=0$, and $U_{2\parallel }\leq U$ when $U_{2\angle }=0$; in
the isotropic case $U_{2\parallel }=U_{2\angle }\equiv U_{2}$,$\ $Eq.~(\ref%
{NNNFM}) is reduced to $U_{2}\leq U/3$.

It is natural to wonder whether the fully polarized ferromagnetic multiplet
remains the ground states of Eq.~(\ref{HintNNN}) at half filling when the
sufficient condition Eq.~(\ref{NNNFM}) is violated. To answer the question
we perform exact diagonalization on Eq.~(\ref{HintNNN}). Assuming a system
size of $L$ unit cells along the edge, the number of different edge state
momenta allowed is approximately$\ N=L/3$. It is convenient to take
advantage of the good quantum numbers of the Hamiltonian, namely the $z$
component of the total spin $S_{z}$ and also the total momentum $Q$ along
the edge direction.\cite{PhysRevLett.101.036803} We measure $Q$ relative to
the fully polarized state $\left\vert \text{FM}\uparrow \right\rangle $
where every edge state is singly occupied by a spin-up electron; for this
state $S_{z}=N/2$ and $Q=0$.

\begin{figure}[ptb]
\includegraphics[width=0.6\textwidth]{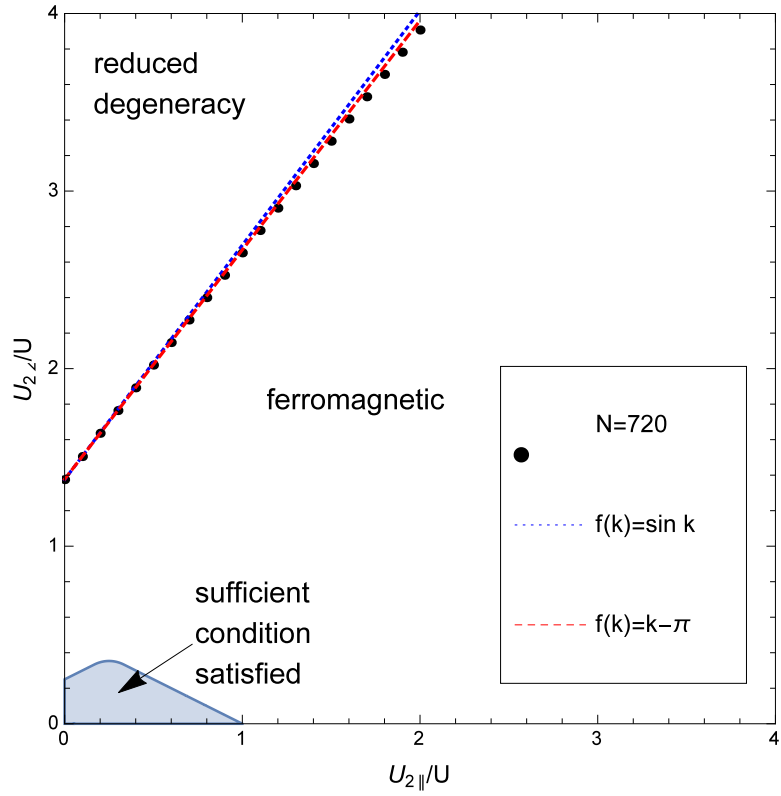}%
\caption{The ground state
phase diagram of Eq.~(\ref{HintNNN}) at half filling on the
$U_{2\parallel}/U$-$U_{2\angle}/U$ plane. The ground states are ferromagnetic below the phase boundary, and have reduced degeneracy above the boundary. The boundary is obtained by exact diagonalization in a system with $N=720$ in the $S_{z}=N/2-1$ sector, and is well approximated by the two straight lines corresponding to the trial wave functions $f\left(k\right) \propto \sin k$ and $f\left(k\right) \propto k-\pi$ (see text). Also shown is the much smaller region where the sufficient condition for ferromagnetic ground states, Eq.~(\ref{NNNFM}), is satisfied.\label{fig:NNNFM}}
\end{figure}

In Fig.~\ref{fig:NNNFM} we plot the ferromagnetic phase boundary for Eq.~(%
\ref{HintNNN}) on the $U_{2\parallel }$-$U_{2\angle }$\ plane, obtained from
exact diagonalization. For comparison we also show the region where the
sufficient condition Eq.~(\ref{NNNFM}) is satisfied. In most of the
parameter space, we find that the ground states at half filling are uniquely
given by the $\left( N+1\right) $-fold degenerate ferromagnetic multiplet
with $S_{z}=-N/2$, $-N/2+1$, ..., $N/2$, and $Q=0$. In particular, the
ground states are always ferromagnetic in the isotropic case $U_{2\parallel
}=U_{2\angle }$. However, in the region above the phase boundary where $%
U_{2\angle }$ is relatively large compared to both $U$ and $U_{2\parallel }$%
, the ground states are not part of the ferromagnetic multiplet, but rather
form a negative-energy manifold with a lower degeneracy and a lower total
spin. For fixed $U$ and $U_{2\parallel }$, the degeneracy is reduced as $%
U_{2\angle }$ gradually increases, and eventually for sufficiently large $%
U_{2\angle }$ the ground state becomes a non-degenerate singlet state in the 
$S_{z}=0$ sector.

\begin{figure}[ptb]
\includegraphics[width=0.8\textwidth]{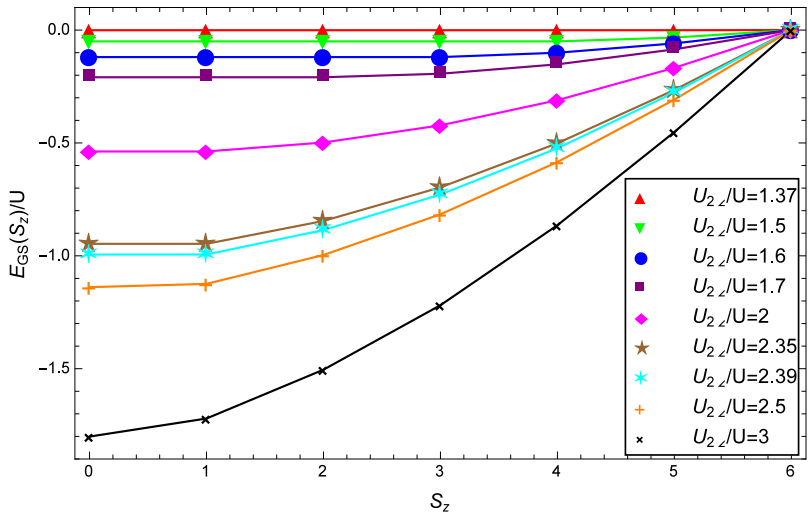} \caption{The ground state
energy in the $S_{z}$ sector, $E_{GS}\left( S_{z}\right)$, versus
$\left\vert S_{z}\right\vert $ for $U_{2\parallel }=0$ and various
$U_{2\angle }/U$ outside of the ferromagnetic regime. The results are
obtained by exact diagonalization in a system with
$N=12$.\label{fig:NNNGSED}}
\end{figure}

In Fig.~\ref{fig:NNNGSED}, choosing a fixed $U_{2\parallel }/U$, we plot $E_{%
\text{GS}}\left( S_{z}\right) $ (the ground state energy in the sector
labeled by $S_{z}$) as a function of $\left\vert S_{z}\right\vert $ for
different $U_{2\angle }/U$ outside of the ferromagnetic regime. We observe
that $E_{\text{GS}}\left( S_{z}\right) $ is a monotonically increasing
function of $\left\vert S_{z}\right\vert $ in general, and becomes a
strictly increasing function of $\left\vert S_{z}\right\vert $ if $%
U_{2\angle }\ $is sufficiently large. This property of $E_{\text{GS}}\left(
S_{z}\right) $ allows us to determine the ferromagnetic phase boundary in
Fig.~\ref{fig:NNNFM} by calculating $E_{\text{GS}}\left( S_{z}=N/2-1\right) $%
, which for a given $N$ is considerably less numerically intensive than $E_{%
\text{GS}}\left( S_{z}=0\right) $. Reasonably accurate estimates of the
phase boundary can then be made through a variational calculation. We can
characterize an arbitrary $Q=0$ state in the $S_{z}=N/2-1$ sector by

\begin{equation}
\int_{\frac{2\pi }{3}}^{\frac{4\pi }{3}}dkf\left( k\right) e_{k,\uparrow
}e_{k,\downarrow }^{\dag }\left\vert \text{FM}\uparrow \right\rangle \text{.}
\end{equation}%
The ferromagnetic state in this sector corresponds to $f\left( k\right) =1$,
i.e. an equal-weighted superposition of all states where every edge state is
singly occupied. The energy expectation value as a functional of $f$ is a
linear combination of $U$, $U_{2\parallel }$ and $U_{2\angle }$:

\begin{equation}
E\left[ f\right] =UC_{0}\left[ f\right] +U_{2\parallel }C_{2\parallel }\left[
f\right] +U_{2\angle }C_{2\angle }\left[ f\right] \text{.}
\end{equation}%
If $E\left[ f\right] <0$, the ground states cannot be the ferromagnetic
multiplet whose energy is always zero. For $f\left( k\right) \propto \sin k$%
, $C_{0}=0.100$, $C_{2\parallel }=0.0964$ and $C_{2\angle }=-0.0730$; for $%
f\left( k\right) \propto k-\pi $, $C_{0}=0.0946$, $C_{2\parallel }=0.0887$
and $C_{2\angle }=-0.0687$. For these two trial wave functions, the
trajectories above which $E\left[ f\right] <0$ are plotted in Fig.~\ref%
{fig:NNNFM}; both trajectories are very close to the ferromagnetic phase
boundary obtained from exact diagonalization.

It should also be cautioned that anisotropy is not necessary to stabilize
non-ferromagnetic ground states. For instance, we can also study an
isotropic interaction consisting of an on-site term and six
fifth-nearest-neighbor terms (or equivalently, next-nearest-neighbor terms
on the same sublattice):

\begin{equation}
U_{\left( 0,0\right) }\equiv U\text{, }U_{\left( 0,\pm 2\right) }=U_{\left(
\pm 3,1\right) }=U_{\left( \pm 3,-1\right) }\equiv U_{5}\text{,}
\end{equation}%
and $U_{\left( \delta _{m},\delta _{n}\right) }=0$ for other $\left( \delta
_{m},\delta _{n}\right) $. For this model

\begin{equation}
\tilde{U}\left( K,q\right) =U+2U_{5}\left( \cos 2K+2\cos K\cos \frac{3q}{2}%
\right) \text{,}
\end{equation}%
so our sufficient condition for ferromagnetism becomes $U_{5}\leq U/3$. In a
system with $N=720$, exact diagonalization shows that a non-ferromagnetic
ground state appears when $U_{5}>80.48U$, i.e. when the non-local $U_{5}$
term is far stronger than the on-site interaction.

Our exact diagonalization results for both models indicate that while
ferromagnetism is favored by the on-site interaction, it may be destabilized
by sufficiently strong non-local interactions. This is in agreement with the
findings of Ref.~\onlinecite{PhysRevLett.111.036601} that the effective
on-site part of the interaction in bulk graphene is reduced by a weighted
average of non-local interactions.

\begin{figure}[ptb]
\includegraphics[width=0.6\textwidth]{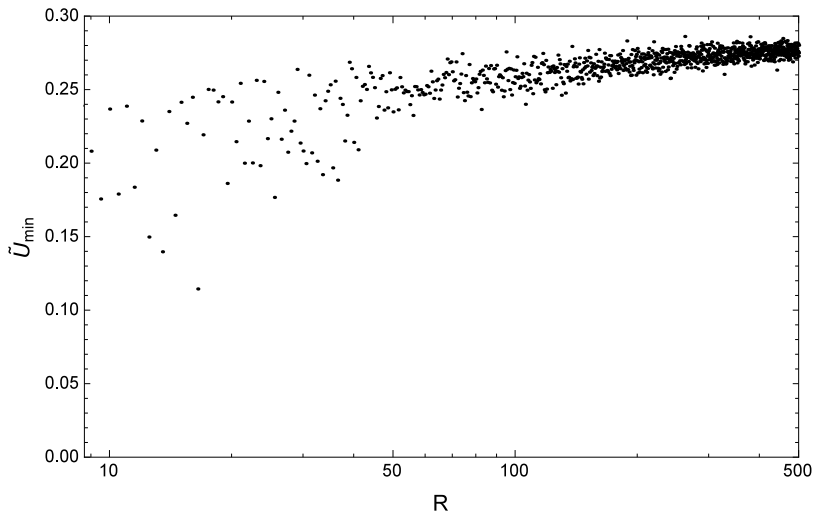}\caption{The minimum of
$\tilde{U}\left( K,q\right) $ for $0\leq K\leq \pi $ and $-2\pi /3\leq q\leq
2\pi /3$, $\tilde{U}_{\min }$, versus $R$, the long-distance cutoff
introduced artificially in the Coulomb interaction
Eq.~(\ref{CoulombU}).\label{fig:CoulombFM}}
\end{figure}

We now investigate whether the unscreened Coulomb interaction, Eq.~(\ref%
{CoulombU}), satisfies the sufficient condition for ferromagnetism. To this
end, we introduce a long-distance cutoff $R$, and minimize $\tilde{U}\left(
K,q\right) $ for the interaction that is given by Eq.~(\ref{CoulombU}) for $%
\left\vert \vec{\delta}\right\vert \leq R$ but vanishes for $\left\vert \vec{%
\delta}\right\vert >R$. In Fig.~\ref{fig:CoulombFM} we show $\tilde{U}_{\min
}$, the minimum of $\tilde{U}\left( K,q\right) $ for $0\leq K\leq \pi $ and $%
-2\pi /3\leq q\leq 2\pi /3$, as a function of $R$ for $R\leq 500$. While $%
\tilde{U}_{\min }$ oscillates wildly, its lower envelope is an increasing
function of $R$, and $\tilde{U}_{\min }$ does not go below $0.2U_{0}$ for $%
50\leq R\leq 500$. This strongly implies that $\tilde{U}_{\min }$ remains
positive as $R\rightarrow \infty $,\ and provides evidence that the
ferromagnetic multiplet states are the unique ground states for the
unscreened Coulomb interaction.

A remark is in order about the short-distance cutoff $d=1/\left( 2\sqrt{3}%
\right) $ in Eq.~(\ref{CoulombU}). If $d$ is treated as a tunable parameter
of our model, then the observation that $\tilde{U}_{\min }\left(
R\rightarrow \infty \right) >0$\ is only valid when $d\lesssim 1$. If $d$ is
close to $1$, $\tilde{U}_{\min }$ oscillates around zero even for $R$ up to $%
500$. Nevertheless, as shown in the next-nearest-neighbor model and the
fifth-nearest-neighbor model, violation of the sufficient condition for
ferromagnetism $\tilde{U}_{\min }\geq 0$ is not an indication of ground
states being non-ferromagnetic. Indeed, we have verified in the $S_{z}=N/2-1$
sector that the ground states remain ferromagnetic for $R$ up to $20$ and $d$
up to $10$.

\section{Low-energy excitations\label{sec:exc}}

In this section we discuss the low-energy single-particle, single-hole and
particle-hole excitations of the ferromagnetic ground state, and also the
effect of the particle-hole symmetry breaking term $\Delta $.

It is simplest to consider the excitations from the maximum $S_{z}$ state $%
\left\vert \text{FM}\uparrow \right\rangle $. We can rewrite the projected
Hamiltonian of Eq.~(\ref{Hproj}) in a form which explicitly annihilates $%
\left\vert \text{FM}\uparrow \right\rangle $:

\begin{eqnarray}
H_{\text{int}} &=&\int dk\left[ \epsilon _{p}\left( k\right) e_{k,\downarrow
}^{\dag }e_{k,\downarrow }+\epsilon _{h}\left( k\right) e_{k,\uparrow
}e_{k,\uparrow }^{\dag }\right] -\int \frac{dkdk^{\prime }dq}{2\pi }\Gamma
\left( k,k^{\prime },q\right) e_{k,\uparrow }e_{k^{\prime }-q,\downarrow
}^{\dag }e_{k^{\prime },\downarrow }e_{k+q,\uparrow }^{\dag }  \notag \\
&&+\frac{1}{2}\int \frac{dkdk^{\prime }dq}{2\pi }\Gamma \left( k,k^{\prime
},q\right) \left( e_{k+q,\downarrow }^{\dag }e_{k^{\prime }-q,\downarrow
}^{\dag }e_{k^{\prime },\downarrow }e_{k,\downarrow }+e_{k+q,\uparrow
}e_{k^{\prime }-q,\uparrow }e_{k^{\prime },\uparrow }^{\dag }e_{k,\uparrow
}^{\dag }\right) \text{,}  \label{Hintprojph}
\end{eqnarray}%
where the domains of integration are such that all edge states have momenta
between $2\pi /3$ and $4\pi /3$, the interaction kernel is

\begin{eqnarray}
&&\Gamma \left( k,k^{\prime },q\right)   \notag \\
&=&\frac{g_{0}\left( k\right) g_{0}\left( k^{\prime }\right) g_{0}\left(
k+q\right) g_{0}\left( k^{\prime }-q\right) }{1-16\cos \frac{k}{2}\cos \frac{%
k+q}{2}\cos \frac{k^{\prime }}{2}\cos \frac{k^{\prime }-q}{2}}\frac{1}{2}%
\sum_{\delta _{m},\delta _{n}}U_{\left( \delta _{m},\delta _{n}\right) }\cos 
\frac{q\delta _{m}}{2}  \notag \\
&&\times \left[ \left( 4\cos \frac{k^{\prime }}{2}\cos \frac{k^{\prime }-q}{2%
}\right) ^{\left\vert \delta _{n}\right\vert }+\left( 4\cos \frac{k}{2}\cos 
\frac{k+q}{2}\right) ^{\left\vert \delta _{n}\right\vert }\right] \text{,}
\label{Gamma}
\end{eqnarray}%
and the energy to create one single spin-down electron or one single spin-up
hole is

\begin{equation}
\epsilon _{p/h}\left( k\right) =\frac{1}{2}\int_{\frac{2\pi }{3}}^{\frac{%
4\pi }{3}}\frac{dk^{\prime }}{2\pi }\Gamma \left( k,k^{\prime },k^{\prime
}-k\right) \pm \Delta \left( 2\cos k+1\right) \text{.}  \label{epsilonph}
\end{equation}%
As noted in Refs.~\onlinecite{PhysRevB.83.195432,PhysRevB.86.115446}, the
interaction $\Gamma \left( k,k^{\prime },q\right) $ is strongly
momentum-dependent. For both Hubbard and Coulomb interactions, $\Gamma
\left( k,k^{\prime },q\right) $ is positive so that spin-down electrons
attract spin-down holes, which favors the formation of bound states between
the two. The third term in Eq.~(\ref{Hintprojph}) generally gives rise to
interaction between edge states with the same spin orientation, although for
the Hubbard model it vanishes due to an additional symmetry of the kernel, $%
\Gamma \left( k,k^{\prime },q\right) =\Gamma \left( k,k^{\prime },k^{\prime
}-k-q\right) $.

\subsection{Single particle and single hole excitations}

We first examine the eigenstates deviating slightly from half-filling,
namely the single particle excitations and single hole excitations. They are
represented by $e_{k,\downarrow }^{\dag }\left\vert \text{FM}\uparrow
\right\rangle $ [of energy $\epsilon _{p}\left( k\right) $] and $%
e_{k,\uparrow }\left\vert \text{FM}\uparrow \right\rangle $ [of energy $%
\epsilon _{h}\left( k\right) $] respectively. Using the definitions Eqs.~(%
\ref{epsilonph}) and (\ref{Gamma}) and the fact that $\delta _{n}+\delta _{m}
$ is even, it is easy to show that $\epsilon _{p/h}\left( k\right) =\epsilon
_{p/h}\left( 2\pi -k\right) $, so we may focus on $2\pi /3\leq k\leq \pi $.

Near the Dirac point\ $0<k-2\pi /3\ll 1$, we can expand Eq.~(\ref{epsilonph}%
) to obtain

\begin{equation}
\epsilon _{p/h}\left( k\right) \approx \left( v\mp \sqrt{3}\Delta \right)
\left( k-\frac{2\pi }{3}\right) \text{,}  \label{epsilonphDir}
\end{equation}%
where the velocity $v$ depends only on the interactions:

\begin{equation}
v\equiv \frac{\sqrt{3}}{2}\sum_{\delta _{m},\delta _{n}}U_{\left( \delta
_{m},\delta _{n}\right) }\int_{\frac{2\pi }{3}}^{\frac{4\pi }{3}}\frac{%
dk^{\prime }}{2\pi }\left( 2\cos \frac{k^{\prime }}{2}\right) ^{\left\vert
\delta _{n}\right\vert }\cos \left( k^{\prime }-\frac{2\pi }{3}\right) \frac{%
\delta _{m}}{2}\text{.}  \label{vFermi}
\end{equation}%
Since the $k^{\prime }$ integral is finite, $v$ is finite for any
short-range interaction. Eq.~(\ref{epsilonphDir}) shows that, as in the
projected Hubbard model,\cite{PhysRevB.86.115446} the single-particle and
single-hole excitations are generally gapless at the Dirac points for a
single zigzag edge.

For the next-nearest-neighbor model Eq.~(\ref{HintNNN}), $v$ is always
positive:

\begin{equation}
v=\frac{\sqrt{3}}{6}U+\frac{3}{4\pi }U_{2\parallel }+\left( \frac{1}{\sqrt{3}%
}-\frac{3}{2\pi }\right) U_{2\angle }\text{.}
\end{equation}%
Nevertheless, $v$ may become negative for certain strongly non-local
interactions. An example is the term with $\delta _{m}=4$ and $\delta _{n}=0$%
, which gives a coefficient of $-3/\left( 16\pi \right) $. The ferromagnetic
ground state will be unstable against the creation of electrons or holes
near the Dirac points in the case of $v<0$, or more generally $v<\sqrt{3}%
\left\vert \Delta \right\vert $ when the particle-hole symmetry breaking
term $\Delta $ is nonzero.

The case of unscreened Coulomb interaction Eq.~(\ref{CoulombU}) is
especially interesting. In this case the low-energy behavior of $\epsilon
_{p/h}\left( k\right) $ is controlled by the long range part of $U_{\left(
\delta _{m},\delta _{n}\right) }$. When $\left\vert \delta _{n}\right\vert
\gg 1$ or $\left\vert \delta _{m}\right\vert \gg 1$, the $k^{\prime }$
integral is dominated by $k^{\prime }$ near the Dirac points, and we find%
\cite{PhysRevLett.96.036801}

\begin{eqnarray}
&&\int_{\frac{2\pi }{3}}^{\frac{4\pi }{3}}\frac{dk^{\prime }}{2\pi }\left(
2\cos \frac{k^{\prime }}{2}\right) ^{\left\vert \delta _{n}\right\vert }\cos
\left( k^{\prime }-\frac{2\pi }{3}\right) \frac{\delta _{m}}{2}  \notag \\
&\approx &\frac{1}{2\pi }\operatorname{Re}\left[ \frac{2}{\sqrt{3}\left\vert \delta
_{n}\right\vert -i\delta _{m}}+\left( -1\right) ^{\left\vert \delta
_{n}\right\vert }e^{i\frac{\pi }{3}\delta _{m}}\frac{2}{\sqrt{3}\left\vert
\delta _{n}\right\vert +i\delta _{m}}\right] \text{.}
\end{eqnarray}%
Approximating the sum over $\delta _{n}$ and $\delta _{m}$ by integrals over 
$x=\delta _{m}/2$ and $y=\sqrt{3}\delta _{n}/2$, and discarding the
subleading contribution from the oscillating term, we see that $v\propto \ln
R$ where $R$ is the long-distance cutoff:

\begin{eqnarray}
v &\approx &\int dx\int dy\frac{U_{0}d}{\sqrt{x^{2}+y^{2}}}\frac{1}{2\pi }%
\frac{\left\vert y\right\vert }{x^{2}+y^{2}}  \notag \\
&\approx &\frac{U_{0}d}{2\pi }\int_{0}^{2\pi }d\theta \left\vert \sin \theta
\right\vert \int_{d}^{R}\frac{dr}{r}=\frac{2U_{0}d}{\pi }\ln \frac{R}{d}%
\text{.}
\end{eqnarray}%
As $R\rightarrow \infty $, the only other large distance scale in the
problem is given by the inverse distance to the Dirac points, which should
therefore replace $R$ as the distance cutoff. In other words, for the
unscreened Coulomb interaction, $\epsilon _{p/h}$ has the following behavior
for $0<k-2\pi /3\ll 1$:

\begin{equation}
\epsilon _{p/h}\left( k\right) \approx \frac{2U_{0}d}{\pi }\left( k-\frac{%
2\pi }{3}\right) \ln \frac{\Lambda }{k-\frac{2\pi }{3}}\text{,}
\label{epsilonphCouDir}
\end{equation}%
where $\Lambda \ll 1$ is a momentum cutoff. This behavior is not affected by
the particle-hole symmetry breaking term $\Delta $, which merely shifts $%
\Lambda $.

\begin{figure}[ptb]
\includegraphics[width=1\textwidth]{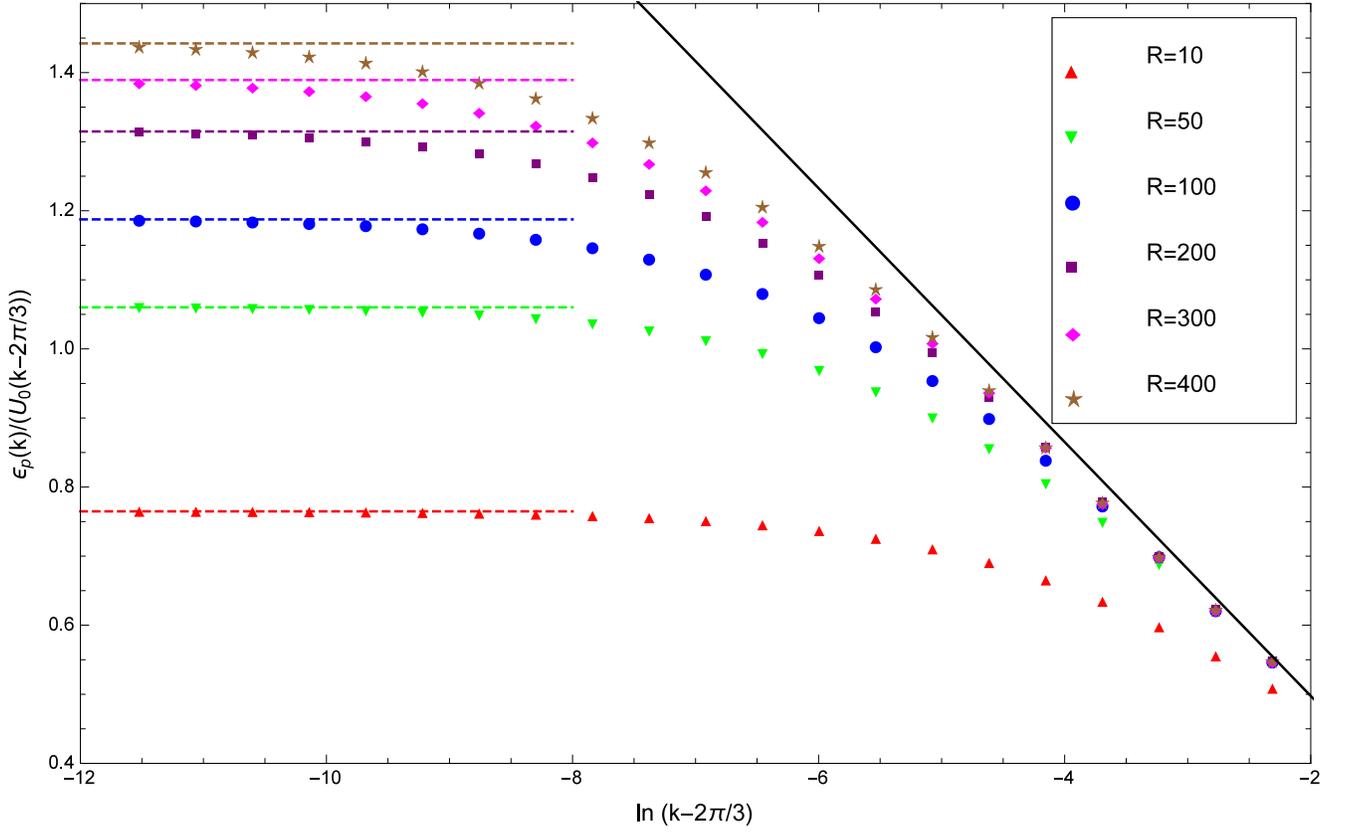}\caption{The
single-particle/single-hole dispersion for the Coulomb interaction near the
Dirac point. $\epsilon _{p/h}\left( k\right) /\left( U_{0} \left( k-2\pi
/3\right) \right)$ is plotted against $\ln \left( k-2\pi /3 \right)$ for
$10^{-5}\leq k-2\pi /3 \leq 0.1$ and different values of long-distance
cutoff $R$, with the particle-hole symmetry breaking perturbation $\Delta$
set to zero. For comparison we also show the velocity given by
Eq.~(\ref{vFermi}) for each $R$ as a horizontal line. The black line has a
slope of $2d/\pi$.\label{fig:CoulombepsilonDir}}
\end{figure}

\begin{figure}[ptb]
\includegraphics[width=0.6\textwidth]{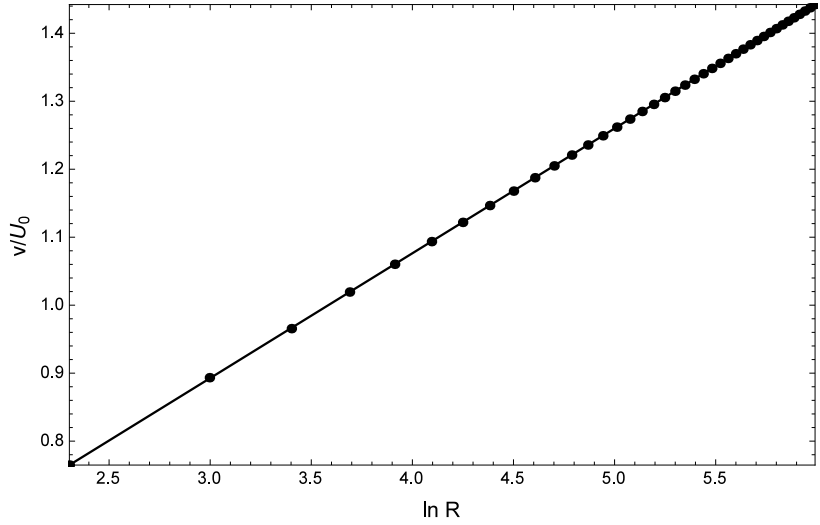}\caption{The velocity
given by Eq.~(\ref{vFermi}) for the Coulomb interaction as a function of the
long-distance cutoff $R$. The fitted line has a slope of $0.1836$ while
$2d/\pi=0.1838$.\label{fig:Coulombv}}
\end{figure}

In Fig.~\ref{fig:CoulombepsilonDir} we plot $\epsilon _{p/h}\left( k\right)
/\left( k-2\pi /3\right) $ versus $\ln \left( k-2\pi /3\right) $ at $%
0<k-2\pi /3\ll 1$ for the Coulomb interaction with various $R$, and show how
the logarithmic divergence in Eq.~(\ref{epsilonphCouDir}) is cut off at low
energies by $R$. We also plot the velocity $v$ given by Eq.~(\ref{vFermi})
as a function of $\ln R$ in Fig.~\ref{fig:Coulombv}. These results suggest
that the Coulomb interaction produces a divergent \textquotedblleft Fermi
velocity\textquotedblright\ for edge modes near the Dirac points, a behavior
reminiscent of the marginal Fermi liquid in bulk graphene with Coulomb
interaction.\cite{PhysRevB.59.R2474}

It is also useful to consider $k=\pi $, since this is where $\epsilon
_{p/h}\left( k\right) $ obtains its maximum for the Hubbard interaction and
the Coulomb interaction, in the absence of particle-hole symmetry breaking.
At $k=\pi $ Eq.~(\ref{epsilonph}) is again greatly simplified:

\begin{eqnarray}
\epsilon _{p/h}\left( \pi \right)  &=&\mp \Delta +\left( \frac{\sqrt{3}}{%
2\pi }-\frac{1}{6}\right) U_{\left( 0,0\right) }+\left( \frac{1}{6}-\frac{%
\sqrt{3}}{8\pi }\right) U_{\left( 1,0\right) }  \notag \\
&&+\frac{1}{\pi }\sum_{\delta _{m}}^{\prime }U_{\left( \delta _{m},0\right) }%
\left[ \frac{8}{\delta _{m}\left( \delta _{m}^{2}-4\right) }\sin \frac{\pi }{%
6}\delta _{m}-\frac{4\sqrt{3}}{\delta _{m}^{2}-4}\cos \frac{\pi }{6}\delta
_{m}\right] \text{,}
\end{eqnarray}%
where the sum is over even $\delta _{m}$ with $\delta _{m}\geq 4$.

$\epsilon _{p/h}\left( \pi \right) $ is also finite for any short-range
interaction. Interestingly, $\epsilon _{p/h}\left( \pi \right) $ depends on $%
U_{\left( \delta _{m},\delta _{n}\right) }$ only if $\delta _{n}=0$: it is,
for instance, independent of $U_{2\angle }$ in the next-nearest-neighbor
model Eq.~(\ref{HintNNN}). For the Coulomb interaction Eq.~(\ref{CoulombU}),
the $\delta _{m}$ sum turns out to be convergent, and we find

\begin{equation}
\epsilon _{p/h}\left( \pi \right) \approx \mp \Delta +0.189U_{0}
\end{equation}%
where $U_{0}$ is the on-site interaction strength. Therefore, when $\Delta
>\Delta _{c}\left( R\rightarrow \infty \right) =0.189U_{0}$ for the
unscreened Coulomb interaction, the maximum $S_{z}$ state $\left\vert \text{%
FM}\uparrow \right\rangle $ becomes unstable towards the creation of a
spin-down electron at $k=\pi $, e.g. by absorption from the bulk. [For
Hubbard interaction with strength $U$, the condition is $\Delta >\Delta
_{c}\left( R=0\right) =\left( \sqrt{3}/\left( 2\pi \right) -1/6\right)
U\approx 0.109U$.]\cite{PhysRevB.86.115446} Similarly, when $\Delta <-\Delta
_{c}\left( R\rightarrow \infty \right) $ there is an instability towards the
creation of a spin-up hole at $k=\pi $.

\begin{figure}[ptb]
\includegraphics[width=1\textwidth]{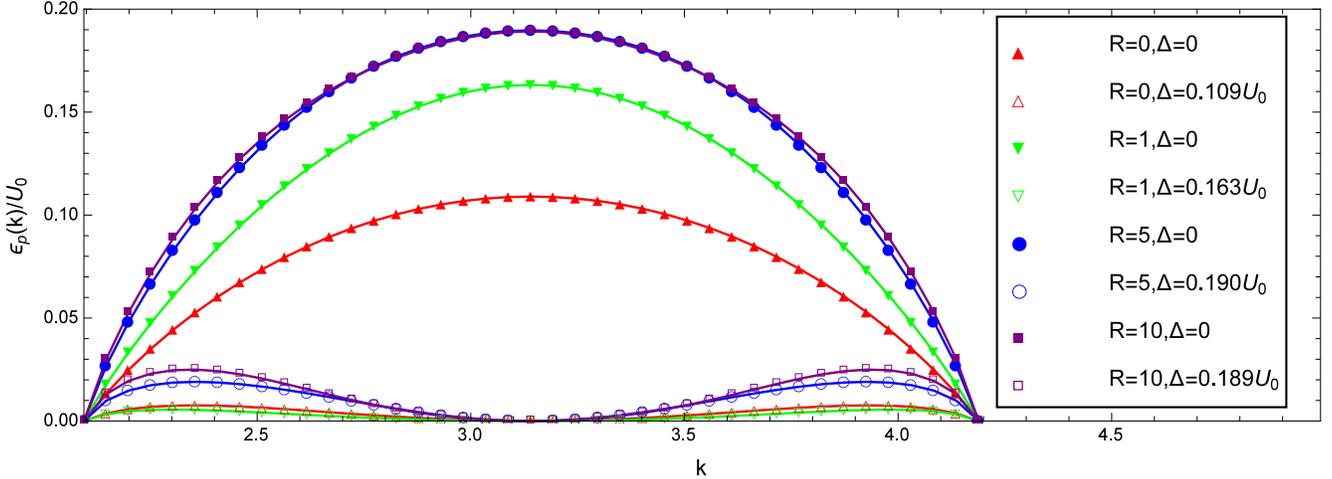}\caption{The
single-particle dispersion for the Coulomb interaction. $\epsilon _{p}\left(
k\right) /U_{0} $ is plotted against $k$ for $2\pi /3 \leq k \leq 4\pi /3$
and different values of long-distance cutoff $R$. The particle-hole symmetry
breaking perturbation $\Delta$ is either $0$ (filled symbols) or
$\Delta_{c}\left( R \right)$ (empty symbols).\label{fig:Coulombepsilon}}
\end{figure}

In Fig.~\ref{fig:Coulombepsilon} we plot $\epsilon _{p}\left( k\right) $
versus $k$ for $2\pi /3\leq k\leq 4\pi /3$ for the Coulomb interaction with
different values of long-distance cutoff $R$, both when $\Delta =0$ and when 
$\Delta =\Delta _{c}\left( R\right) $ so that $\epsilon _{p}\left( \pi
\right) $ vanishes. Notice that for the Coulomb interaction $\epsilon
_{p}\left( k\right) >0$\ for $0<k-2\pi /3\ll 1$ even when $\Delta =\Delta
_{c}\left( R\right) $; that is, as we increase $\left\vert \Delta
\right\vert $, single particle or single hole creation energy becomes negative at $k=\pi $ sooner than it does near the Dirac points.

\subsection{1-particle-1-hole sector}

We turn to the half-filled sector with $N-1$ spin-up electrons and $1$
spin-down electron, so that $S_{z}=N/2-1$. This sector hosts $1$ spin-down
electron and $1$ spin-up hole relative to the $\left\vert \text{FM}\uparrow
\right\rangle $ state, and accommodates the excitations that would be seen
as magnons in an effective spin model.

Let the total momentum relative to $\left\vert \text{FM}\uparrow
\right\rangle $ be $Q$, and without loss of generality we assume $Q\geq 0$.
Denoting an eigenstate by

\begin{equation}
\int_{\frac{2\pi }{3}}^{\frac{4\pi }{3}-Q}dkf\left( k;Q\right) e_{k,\uparrow
}e_{k+Q,\downarrow }^{\dag }\left\vert \text{FM}\uparrow \right\rangle \text{%
,}
\end{equation}%
we obtain the following Schroedinger's equation:

\begin{equation}
\left[ E-\epsilon _{h}\left( k\right) -\epsilon _{p}\left( k+Q\right) \right]
f\left( k;Q\right) =-\int \frac{dk^{\prime }}{2\pi }\Gamma \left(
k,k^{\prime }+Q,k^{\prime }-k\right) f\left( k^{\prime };Q\right) \text{,}
\label{1p1hSchEq}
\end{equation}%
where $E$ is the energy eigenvalue. The ferromagnetic state in the
1-particle-1-hole sector, $f\left( k;Q=0\right) =1$, is obviously a
zero-energy solution.

It is possible for $f\left( k,Q\right) $ to have a $\delta $-function peak
at $k=k_{0}$. In this case the solution to Eq.~(\ref{1p1hSchEq}) is part of
the 1-particle-1-hole continuum, and has an energy $E=\epsilon _{h}\left(
k_{0}\right) +\epsilon _{p}\left( k_{0}+Q\right) $. Another possibility is
having $E<\epsilon _{h}\left( k\right) +\epsilon _{p}\left( k+Q\right) $ for
any $k$, in which case $f\left( k;Q\right) $ does not have any $\delta $%
-function peaks, and the solution is a particle-hole bound state, or an
exciton. Since it reduces $S_{z}$ by $1$, it can also be viewed as a magnon
in an effective spin model.

For any short range interaction, we can show that the exciton energy has the
following $Q\ll 1$ behavior:

\begin{equation}
E\left( Q\right) =\frac{3v}{2\pi }\left( 1-\frac{3\Delta ^{2}}{v^{2}}\right)
Q^{2}\ln \frac{\Lambda ^{\prime }}{Q}\text{,}  \label{excdispersion}
\end{equation}%
where $v$ is the velocity Eq.~(\ref{vFermi}) that also appears in the single
particle dispersion, and $\Lambda ^{\prime }\ll 1$ is again a momentum
cutoff. The inverse exciton mass, or the spin stiffness of the ferromagnetic
zigzag edge, is therefore logarithmically divergent. The derivation of Eq.~(%
\ref{excdispersion}) is sketched in Appendix~\ref{sec:appSchEq}, where we
see the divergence arises due to the linear behavior of $\epsilon
_{p/h}\left( k\right) $ near the Dirac points. This divergence is possibly
related to the large spin stiffness found by Refs.~%
\onlinecite{PhysRevLett.100.047209,PhysRevB.80.155441} for $U$ comparable to 
$t$.

Although similar exciton dispersions have been previously reported in carbon
nanotubes,\cite%
{PhysRevB.75.035407,*PhysRevLett.106.136805,*PhysRevLett.109.187403} in
contrast to Eq.~(\ref{excdispersion}) they originate from the long-range
nature of the Coulomb interaction. In fact, since in the Coulomb interaction
with a long-distance cutoff $R$ we have $v\propto \ln R$, we expect that Eq.
(\ref{excdispersion}) is modified to $E\left( Q\right) \propto Q^{2}\ln
^{2}Q $ for $R\rightarrow \infty $; that is, the spin stiffness is even more
divergent than a logarithm for the unscreened Coulomb interaction. Fig.~\ref%
{fig:CoulombexcEDir} shows $E\left( Q\right) /Q^{2}$ plotted against $\ln Q$
at $0<Q\ll 1$ for some values of $R$ and $\Delta =0$, where $E\left(
Q\right) $ is found by solving Eq.~(\ref{1p1hSchEq}) numerically via
Chebyshev series expansion.\cite{abramowitz2012handbook}

\begin{figure}[ptb]
\includegraphics[width=1\textwidth]{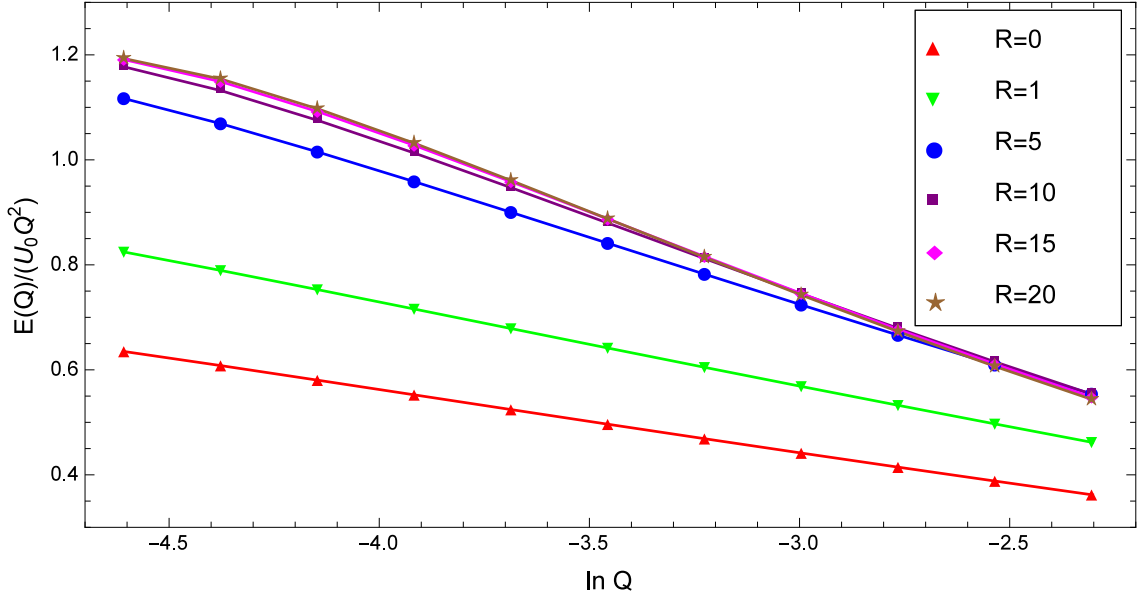}\caption{The exciton
dispersion for the Coulomb interaction at small momenta. $E\left( Q\right)
/\left( U_{0} Q^{2} \right) $ is plotted against $\ln Q$ for $0.01\leq Q
\leq 0.1$ and different values of long-distance cutoff $R$, with the
particle-hole symmetry breaking perturbation $\Delta$ set to zero. The
lowest $100$ Chebyshev polynomials are retained in the numerical
solution.\label{fig:CoulombexcEDir}}
\end{figure}

\begin{figure}[ptb]
\includegraphics[width=1\textwidth]{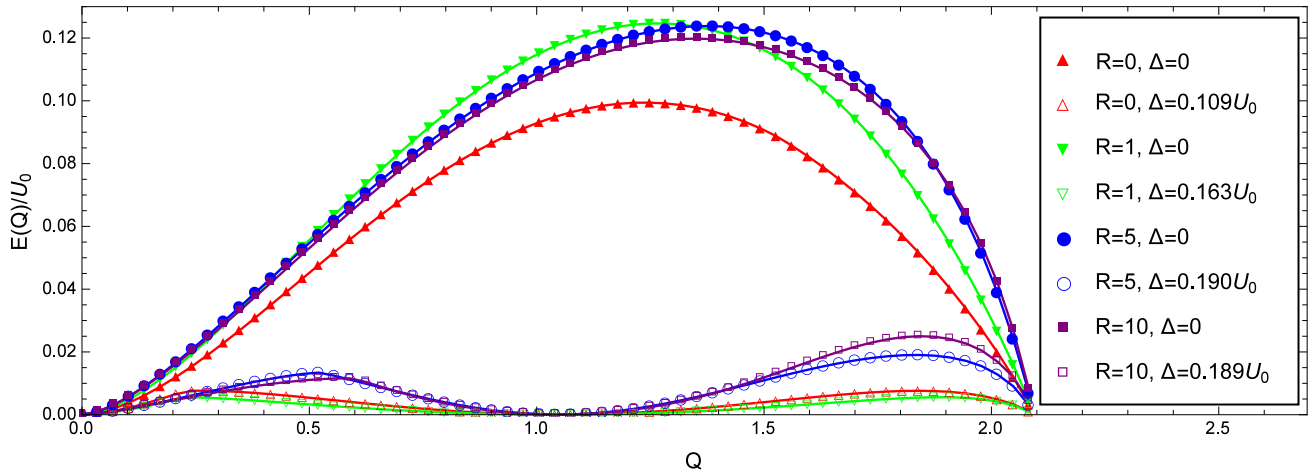}\caption{The exciton
dispersion for the Coulomb interaction. $E\left( Q\right) /U_{0}$ is plotted
against $Q$ for $0 \leq Q \leq 2\pi/3$ and different values of long-distance
cutoff $R$. The particle-hole symmetry breaking perturbation $\Delta$ is
either $0$ (filled symbols) or $\Delta_{c}\left( R \right)$ (empty symbols).
The lowest $100$ Chebyshev polynomials are retained in the numerical
solution.\label{fig:CoulombexcE}}
\end{figure}

It is also helpful to examine the effect of $\Delta $ on the exciton
dispersion, taking as an example the Coulomb interaction with a
long-distance cutoff $R$. As depicted in Fig.~\ref{fig:CoulombexcE}, when $%
\left\vert \Delta \right\vert =\Delta _{c}\left( R\right) $ so that $%
\epsilon _{p}\left( \pi \right) =0$, the exciton dispersion $E\left(
Q\right) $ calculated numerically also approximately vanishes at $Q=\pm \pi
/3$, and the exciton wave function strongly favors the state with a
spin-down electron at $\pi $ and a spin-up hole at either Dirac point. For $%
\left\vert \Delta \right\vert >\Delta _{c}\left( R\right) $, in parallel
with the Hubbard case,\cite{PhysRevB.86.115446} $E\left( \pm \pi /3\right) $
becomes negative which indicates that the ground state at half-filling is no
longer maximally spin polarized; instead, the edge states near $\pi $ become
more likely to be doubly occupied and the edge states near the Dirac points
become more likely to be unoccupied.

Finally, we mention that in the 2-particle-2-hole sector, the excitons in
the 1-particle-1-hole sector can form an additional bound state below the
exciton continuum. Nevertheless, for both the Hubbard and the Coulomb
interactions with $\left\vert \Delta \right\vert <\Delta _{c}\left( R\right) 
$, we find numerically that the bottom of the 2-particle-2-hole bound state
dispersion remains positive; we thus conjecture that the ferromagnetic
ground state is stable for $\left\vert \Delta \right\vert $ up to $\Delta
_{c}\left( R\right) $. We also mention that the bound state picture provides
an intuitive explanation for the non-ferromagnetic regime in Fig.~\ref%
{fig:NNNGSED}: for $M<N/2$, we can usually form an $M$-particle-$M$-hole
bound state with a non-negative binding energy, i.e. with an energy lower
than or equal to the sum of energies of an $\left( M-1\right) $-particle-$%
\left( M-1\right) $-hole bound state and a 1-particle-1-hole bound state.
Therefore, if the 1-particle-1-hole ground state has a negative energy as
happens for sufficiently large $U_{2\angle }$, then as $M$ increases and $%
\left\vert S_{z}\right\vert $ decreases, the ground state energy in the $%
S_{z}$ sector either stays the same or decreases.

\section{Discussion and conclusions\label{sec:outlook}}

In our effective model Eq.~(\ref{Hintproj}) at $O\left( U\right) $, we have ignored the dynamics of low-lying bulk degrees of freedom near the Dirac points, so an obvious issue is whether this approximation is justified. For the on-site
Hubbard interaction, the answer is partly given by Refs.~%
\onlinecite{PhysRevB.86.115446,PhysRevB.92.125416}, where effective
Hamiltonians are found to $O\left( U^{2}/t\right) $ by integrating out the
bulk states and neglecting retardation. While Ref.~%
\onlinecite{PhysRevB.86.115446} finds that the $O\left( U^{2}/t\right) $\
correction to the Hamiltonian has a $q^{2}\ln q$ behavior for small momentum
transfer $q$, such behavior does not necessarily hint at a breakdown of the
perturbation theory, as logarithms also appear at $O\left( U\right) $, e.g.
in the exciton dispersion Eq.~(\ref{excdispersion}). Ref.~%
\onlinecite{PhysRevB.92.125416} further shows that, as far as the effective
spin model is concerned, the interaction strengths are only weakly modified
by the bulk states even for $U$ comparable to $t$. In other words, there is
no evidence that the perturbation theory in $U/t$ is divergent. However,
while a weak Hubbard interaction is known to be irrelevant in the bulk, a
weak Coulomb interaction is marginally irrelevant and may lead to further
logarithmic corrections.\cite{PhysRevB.59.R2474,RevModPhys.84.1067} It
therefore remains an open question whether integrating out the bulk states
at $O\left( U^{2}/t\right) $ qualitatively changes the physics of the $%
O\left( U\right) $ edge model for the unscreened Coulomb interaction.

Another problem that we have not discussed so far is the inter-edge coupling
in realistic graphene nanoribbons. We now consider a ribbon of large but
finite width $W\gg 1$ with two zigzag edges, whose overall ground state is
antiferromagnetic. The inter-edge coupling originates in part from the
direct interaction between opposite edges, which is significant even at the
first order in interaction if it is long-ranged [$O\left( U_{0}/W\right) $
in the Coulomb case]. Inter-edge coupling is also mediated by bulk states,
which is second order in interaction and is $O\left( U^{2}/\left(
tW^{2}\right) \right) $ in the Hubbard case.\cite{PhysRevB.86.115446} Yet
another source is the hopping amplitude between edge states of opposite
edges, which exists even in the absence of interactions and leads to an
energy gap exponentially small in $W$. For wide ribbons $W\gg 1$, it is well
known that the edge states are no longer strictly localized near one edge
when their momenta are within $O\left( 1/W\right) $ of the Dirac points. The
hopping amplitude at momentum $k$ thus grows rapidly as $k$ approaches the
Dirac points, eventually reaching $O\left( t/W\right) $.\cite%
{PhysRevB.73.235411,PhysRevLett.102.227205} Under our assumption $U\ll t$,
this is actually a much larger energy scale than that of the direct
inter-edge Coulomb interaction or that of the bulk-mediated inter-edge
interaction. Thus it is not justified to ignore the inter-edge hopping
amplitude near the Dirac points in the effective model for a nanoribbon. In
fact, at the mean-field level, it is exactly the part of the Brillouin zone
near the Dirac points that contributes the most to the inter-edge
superexchange interaction,\cite{PhysRevLett.102.227205,PhysRevB.83.165415}
and the spin wave dispersion becomes linear for small momenta once the
inter-edge coupling is taken into account.\cite{1367-2630-13-3-033028}
Although an effective edge model incorporating the inter-edge hopping\cite%
{PhysRevB.87.245431} is often much less analytically accessible beyond the
mean-field level, we hope further insight on the effect of Coulomb
interaction in finite width nanoribbons can be gleaned from exact
diagonalization.

In conclusion, we have investigated the effects of long-range interactions on
the zigzag edge states of a semi-infinite graphene sheet. By projecting the
interaction onto the edge state subspace, we obtain an effective model for
which the states in the maximally polarized ferromagnetic multiplet are zero
energy eigenstates. A sufficient condition is found for the ferromagnetic
multiplet to be the ground states, and we present evidence that the
unscreened Coulomb interaction satisfies this condition, which implies that
its ground states are ferromagnetic. In cases where the sufficient condition
is not met, exact diagonalization results indicate that the ground state can
be non-ferromagnetic, provided that certain non-local components of the
interaction are sufficiently strong. We also discuss the single-particle
excitations, single-hole excitations and spin-1 excitons of the maximum $%
S_{z}$ ground state. For short range interactions the single-particle and
single-hole excitations have linear dispersions near the Dirac points, as
described in Eq.~(\ref{epsilonphDir}). The slope $v$ also governs the
exciton energy at small momenta, Eq.~(\ref{excdispersion}), which shows a $%
vQ^{2}\ln Q$ behavior. For the unscreened Coulomb interaction $v$ becomes
logarithmically divergent as a function of the long-distance cutoff,
corresponding to a $\delta k\ln \delta k$ behavior where $\delta k\ll 1$ is
the distance from either of the Dirac points. The edge states acquire a
dispersion due to a particle-hole symmetry breaking perturbation $\Delta $;
the ferromagnetic ground state can be destroyed if $\left\vert \Delta
\right\vert $ is large enough.

\begin{acknowledgments}
This work was supported in part by NSERC of Canada, Discovery Grant
04033-2016 and the Canadian Institute for Advanced Research. ZS would like
to acknowledge helpful discussions with Emilian Nica.
\end{acknowledgments}

\appendix

\section{Exciton dispersion at small momenta for a single zigzag edge with
short-range interactions\label{sec:appSchEq}}

For simplicity, we illustrate the derivation of Eq.~(\ref{excdispersion})
with the Hubbard interaction $U_{\left( 0,0\right) }=U$. Generalization to
non-local interactions is tedious but straightforward; it is briefly
discussed at the end of this appendix.

Expanding the denominator of the kernel $\Gamma $ in Eq.~(\ref{1p1hSchEq}),
we can isolate the $k$ dependence of $f\left( k;Q\right) $:

\begin{equation}
f\left( k;Q\right) =-\frac{g_{0}\left( k\right) g_{0}\left( k+Q\right) }{%
E-\epsilon _{h}\left( k\right) -\epsilon _{p}\left( k+Q\right) }%
\sum_{l=0}^{\infty }\left( 4\cos \frac{k}{2}\cos \frac{k+Q}{2}\right)
^{l}U\Phi _{l}\left( Q\right) \text{,}  \label{fPhi}
\end{equation}%
where $\Phi $'s are independent of $k$, and are defined as

\begin{equation}
\Phi _{l}\left( Q\right) =\int_{\frac{2\pi }{3}}^{\frac{4\pi }{3}-Q}\frac{%
dk^{\prime }}{2\pi }g_{0}\left( k^{\prime }+Q\right) g_{0}\left( k^{\prime
}\right) \left( 4\cos \frac{k^{\prime }}{2}\cos \frac{k^{\prime }+Q}{2}%
\right) ^{l}f\left( k^{\prime };Q\right) \text{.}  \label{Phif}
\end{equation}%
Inserting Eq.~(\ref{fPhi}) into Eq.~(\ref{Phif}), we obtain an infinite
number of linear equations satisfied by $\Phi $:

\begin{eqnarray}
\Phi _{l}\left( Q\right) &=&-\int \frac{dk^{\prime }}{2\pi }\frac{\left(
1-4\cos ^{2}\frac{k^{\prime }+Q}{2}\right) \left( 1-4\cos ^{2}\frac{%
k^{\prime }}{2}\right) }{E-\epsilon _{h}\left( k^{\prime }\right) -\epsilon
_{p}\left( k^{\prime }+Q\right) }\left( 4\cos \frac{k^{\prime }}{2}\cos 
\frac{k^{\prime }+Q}{2}\right) ^{l}  \notag \\
&&\times \sum_{l^{\prime }=0}^{\infty }\left( 4\cos \frac{k^{\prime }}{2}%
\cos \frac{k^{\prime }+Q}{2}\right) ^{l^{\prime }}U\Phi _{l^{\prime }}\left(
Q\right) \text{.}  \label{PhiPhi}
\end{eqnarray}%
For $Q\ll 1$, the integrand on the right-hand side can be expanded to $%
O\left( E\right) $ and $O\left( Q^{2}\right) $.

\begin{eqnarray}
\Phi _{l}\left( Q\right) &=&\int \frac{dk^{\prime }}{2\pi }\frac{\left(
1-4\cos ^{2}\frac{k^{\prime }}{2}\right) ^{2}}{\epsilon _{h}\left( k^{\prime
}\right) +\epsilon _{p}\left( k^{\prime }\right) }\left( 4\cos ^{2}\frac{%
k^{\prime }}{2}\right) ^{l}\sum_{l^{\prime }=0}^{\infty }\left( 4\cos ^{2}%
\frac{k^{\prime }}{2}\right) ^{l^{\prime }}U\Phi _{l^{\prime }}\left(
Q\right)  \notag \\
&&+E\int \frac{dk^{\prime }}{2\pi }\frac{\left( 1-4\cos ^{2}\frac{k^{\prime }%
}{2}\right) ^{2}}{\left[ \epsilon _{h}\left( k^{\prime }\right) +\epsilon
_{p}\left( k^{\prime }\right) \right] ^{2}}\left( 4\cos ^{2}\frac{k^{\prime }%
}{2}\right) ^{l}\sum_{l^{\prime }=0}^{\infty }\left( 4\cos ^{2}\frac{%
k^{\prime }}{2}\right) ^{l^{\prime }}U\Phi _{l^{\prime }}\left( 0\right) 
\notag \\
&&+2\int_{\frac{2\pi }{3}}^{\frac{2\pi }{3}+\Lambda }\frac{dk^{\prime }}{%
2\pi }\frac{-\frac{3}{4}\left( 1-\frac{3\Delta ^{2}}{v^{2}}\right) Q^{2}}{%
v\left( 2k+Q-\frac{4\pi }{3}\right) -\sqrt{3}\Delta Q}\sum_{l^{\prime
}=0}^{\infty }U\Phi _{l^{\prime }}\left( 0\right) \text{.}
\end{eqnarray}%
In the second and the third lines we have approximated $\Phi _{l^{\prime
}}\left( Q\right) \approx \Phi _{l^{\prime }}\left( 0\right) $, assuming
that $\Phi _{l}\left( Q\right) $ is well-behaved at $Q=0$ and any difference
is $O\left( Q\right) $. In the third line we have retained the most singular
contribution at $O\left( Q^{2}\right) $, which are from the vicinity of the
Dirac points (hence the factor of $2$), as the remaining terms contain no
infrared divergence.

Using Eq.~(\ref{Phif}) and recalling that the $Q=0$ solution is $f\left(
k;0\right) =1$, we have

\begin{equation}
\sum_{l^{\prime }=0}^{\infty }U\Phi _{l^{\prime }}\left( 0\right)
=\sum_{l^{\prime }=0}^{\infty }\int_{\frac{2\pi }{3}}^{\frac{4\pi }{3}}\frac{%
dk}{2\pi }g_{0}^{2}\left( k\right) U\left( 4\cos ^{2}\frac{k}{2}\right)
^{l^{\prime }}=\frac{U}{3}\text{,}  \label{sum0}
\end{equation}%
and

\begin{equation}
\sum_{l^{\prime }=0}^{\infty }\left( 4\cos ^{2}\frac{k^{\prime }}{2}\right)
^{l^{\prime }}U\Phi _{l^{\prime }}\left( 0\right) =\sum_{l^{\prime
}=0}^{\infty }\int_{\frac{2\pi }{3}}^{\frac{4\pi }{3}}\frac{dk}{2\pi }%
g_{0}^{2}\left( k\right) U\left( 16\cos ^{2}\frac{k}{2}\cos ^{2}\frac{%
k^{\prime }}{2}\right) ^{l^{\prime }}=\frac{\epsilon _{h}\left( k^{\prime
}\right) +\epsilon _{p}\left( k^{\prime }\right) }{1-4\cos ^{2}\frac{%
k^{\prime }}{2}}\text{;}  \label{sum1}
\end{equation}%
therefore

\begin{eqnarray}
\Phi _{l}\left( Q\right) &=&\int \frac{dk^{\prime }}{2\pi }\frac{\left(
1-4\cos ^{2}\frac{k^{\prime }}{2}\right) ^{2}}{\epsilon _{h}\left( k^{\prime
}\right) +\epsilon _{p}\left( k^{\prime }\right) }\left( 4\cos ^{2}\frac{%
k^{\prime }}{2}\right) ^{l}\sum_{l^{\prime }=0}^{\infty }\left( 4\cos ^{2}%
\frac{k^{\prime }}{2}\right) ^{l^{\prime }}U\Phi _{l^{\prime }}\left(
Q\right)  \notag \\
&&+E\int \frac{dk^{\prime }}{2\pi }\frac{1-4\cos ^{2}\frac{k^{\prime }}{2}}{%
\epsilon _{h}\left( k^{\prime }\right) +\epsilon _{p}\left( k^{\prime
}\right) }\left( 4\cos ^{2}\frac{k^{\prime }}{2}\right) ^{l}-\frac{1}{8\pi v}%
\left( 1-\frac{3\Delta ^{2}}{v^{2}}\right) Q^{2}U\ln \frac{\Lambda ^{\prime }%
}{Q}\text{.}  \label{PhiPhiTaylor}
\end{eqnarray}%
Now, we multiply the entire expression by $\left[ 1-4\cos ^{2}\left(
k/2\right) \right] \left[ 4\cos ^{2}\left( k/2\right) \right] ^{l}U$, then
sum over $l$ and integrate over $k$. The left hand side then cancels the
first term on the right hand side, and using $v=U/\left( 2\sqrt{3}\right) $,
we are left with Eq.~(\ref{excdispersion}).

In the presence of non-local interactions, one needs to assign three more
indices to $\Phi $, namely $\delta _{m}$, $\delta _{n}$ and $\alpha =1$, $2$
[corresponding to the two terms in the third line of Eq.~(\ref{Gamma})]. All
three indices should be summed over in Eq.~(\ref{PhiPhi}), and subsequently
in Eqs.~(\ref{sum0}), (\ref{sum1}) and (\ref{PhiPhiTaylor}).

\bibliography{CoulombZigzag}

\end{document}